\documentclass[aps,floatfix,prd,showpacs]{revtex4}
\usepackage{graphicx}
\usepackage{dcolumn}
\usepackage{bm}

\begin{document}

\def\be{\begin{equation}}
\def\ee{\end{equation}}
\def\lr{\leftrightarrow}

\title{Parity Symmetry in QED3}
\author{Pok Man Lo and Eric S. Swanson}
\affiliation{
Department of Physics and Astronomy, 
University of Pittsburgh, 
Pittsburgh, PA 15260, 
USA.}

\date{\today}

\begin{abstract}
Schwinger-Dyson equations are used to study 
spontaneous chiral and parity symmetry breaking 
of three dimensional quantum electrodynamics with two-component fermions.
This theory admits a topological photon mass that explicitly breaks parity symmetry and generates a fermion mass. We show that the pattern of symmetry breaking maintains parity but breaks chiral symmetry. We also find that chiral symmetry is restored at a critical number of fermion flavours in our truncation scheme. The Coleman-Hill theorem is used to demonstrate that the results are reasonably accurate.
\end{abstract}

\maketitle

\section{Introduction}

A variety of novel features has spurred interest in low-dimensional QED for many decades\cite{schwing}.
 For example, high temperature QCD can be represented as the dimensionally reduced QCD3. If the number of quark flavours ($N_f$) is large, the nonabelian behaviour of the theory is suppressed and it may be approximated as quantum electrodynamics in three dimensions (QED3)\cite{pis}. Massless QED3 in the large $N_f$ limit generates dynamical fermion masses that are suppressed exponentially in the fermion number. Thus this theory illustrates how large mass hierarchies can be dynamically generated\cite{Appelquist}, which is of interest to BSM physics. 

More recently, QED3 has been used as a model field theory for three dimensional condensed matter systems. Examples include applications to high $T_c$ superconductors, where the relevant 
dynamics is thought to be isolated to copper-oxygen planes in the cuprate\cite{highTc}. It is also considered as a gauge formulation of the 2+1 dimensional Heisenberg spin model\cite{richert}, a possible model for graphene\cite{graph}, and quantum versions of spin-ice\cite{spin-ice}.

When coupled to $N_f$ massless four-component fermions, QED3 exhibits a $U(2N_f)$ symmetry that can be broken to $U(N_f)\times U(N_f)$, which is the three dimensional analogue of chiral symmetry breaking. It is widely agreed that symmetry breaking occurs for low $N_f$. More interesting is the behaviour as the number of fermions becomes large where it is possible that the interaction becomes sufficiently screened that vacuum condensation no longer occurs. Indeed, Appelquist {\it et al.} have used the large $N_f$ limit with additional approximations to argue that there is a critical number of flavours, $N_\star \approx 3.5$ above which the theory remains in the symmetric phase. 
Furthermore, the lack of massless scalar bound states at the critical coupling has been used to argue that the chiral restoration phase transition is not second order, but is of a novel type\cite{jklt}.

The belief that a critical number of fermion flavours exists is not without controversy. For example, Pennington and Walsh assert that wavefunction renormalisation (which is null as $N_f \to \infty$) is central to determining the details of symmetry breaking. Their solution of a truncation of the Schwinger-Dyson equations reveals that, in fact, chiral symmetry is never restored\cite{PW}. In addition, Pisarski has used the renormalisation group to argue that fermion mass generation occurs for all values of $N_f$\cite{p2}. In spite of these claims, more recent Schwinger-Dyson computations again find a critical value of $N_f$ with $N_\star \approx 3.5$\cite{maris,FADM}.

In principle lattice gauge theory can determine the nonperturbative properties of QED3. In practice, lattice computations have been hampered by the large ratio of the coupling and dynamical mass scales.
Compounding the difficulty is the long range nature of the force, which leads to large finite volume effects. Nevertheless, computations with large lattices have been made and find $N_\star \approx 1.5$\cite{latt,sk}. It appears that the discrepancy is due to the necessity of extremely large lattices to achieve the continuum limit. This has been explicitly demonstrated by
Schwinger-Dyson calculations with QED3 on a torus, which also show that quantities can appear to falsely converge at smaller lattice sizes\cite{latt2}.

A novel feature of QED3 is that it is possible to introduce a topological Chern-Simons-like photon mass term to the theory\cite{originals}. This term breaks parity and time reversal symmetries. It is also possible to formulate the theory with two-component fermions.  In this case a nonzero photon mass induces a finite fermion mass at one-loop (and vice versa)\cite{DJT,Appelquist-parity}. This raises the interesting possibility that parity symmetry can be  spontaneously broken in the massless theory. This question was first examined by Appelquist {\it et al.} many years ago\cite{Appelquist-parity}. They concluded that finite fermion masses were dynamically generated, but that these masses appear in pairs of opposite sign, thereby maintaining the parity symmetry of the vacuum and a massless photon, in agreement with a 
general argument of Vafa and Witten\cite{vafa}.  Their conclusions were based on an analytic examination of the Schwinger-Dyson equations in rainbow-ladder approximation. In  addition the authors assumed
the large $N_f$ limit, that there is no fermion wavefunction renormalisation, that the fermion self energy is constant at low momentum, and that it is valid to truncate integrals at the scale $e^2$. Their general conclusions were 
confirmed by other groups with similar methods\cite{stam}. Nevertheless, a surprising counter claim exists, namely
Hoshino and Matsuyma\cite{Hoshino} obtained a nontrivial parity violating numerical solution to the Schwinger-Dyson equations for $N_f=1$. This analysis employed the rainbow-ladder approximation and a bare photon propagator.  There appears to be, furthermore, a community that believes that spontaneous parity symmetry breaking is viable\cite{niemi}.

We revisit the massless two-component theory in an attempt to resolve the discrepancy mentioned above and to examine the robustness of the parity symmetry preserving scenario under a variety of truncation schemes. Since the two-component theory in a reflection invariant vacuum is equivalent to the four-component theory with $N_f/2$ fermions, we can also study the possible existence and stability of the chiral symmetry restoration transition.
Of course one generally seeks stability with respect to truncation variation as a sign of robustness. Unfortunately, this can be difficult to assess in a rather small truncation model space. Gauge invariance is also a useful metric if suitable observables can be found. 
The situation is alleviated somewhat by a theorem due to Coleman and Hill, which gives the exact result for the parity-violating portion of the photon self energy\cite{CH}.  We will use this to check the efficacy of various truncations in the following.

Our investigation will utilise the Schwinger-Dyson equations truncated at the two-point function level with various model fermion-photon vertices. Our most sophisticated model will employ the Ball-Chiu longitudinal vertex with the Curtis-Pennington transverse vertex, putting the computation on a par with  the most recent investigations of four-component QED3\cite{FADM}. We will confirm that nontrivial fermion masses are generated in opposite sign pairs, and therefore the vacuum remains reflection invariant. 
Furthermore, we confirm that chiral symmetry is spontaneously broken at low $N_f$ and is restored at a critical number of flavours that depends weakly on the vertex Ansatz. Finally we establish that the Coleman-Hill theorem is satisfied reasonably well and that this agreement improves with the quality of the vertex, lending support to the conclusions presented here.

\section{QED3 and the Schwinger-Dyson Equations}

We study massless abelian gauge theory in three dimensions with two-component fermions. The Lagrangian is taken to be

\be
{\cal L} = -\frac{1}{4}F^2 + \bar \psi(i\rlap{/}\partial + e\rlap{/}A)\psi - \frac{1}{2\xi}(\partial\cdot A)^2.
\ee
The coupling $e^2$ has units of mass and the theory is superrenormalisable. As mentioned in the introduction, a fermion mass term generates a topological photon mass

\be
{\cal L}_{\rm CS} = \mu\frac{1}{4}\epsilon_{\mu\nu\alpha}F^{\mu\nu}A^\alpha
\ee
via radiative corrections. Similarly, a nonzero value for $\mu$ will generate a fermion mass at one loop\cite{originals, DJT}.

The full photon propagator is given by the expression

\be
D_{\mu\nu} = \frac{-i\, (1-\Pi)}{p^2(1-\Pi)^2 - (\mu-\tilde\Pi)^2}\,\left( P_{\mu\nu} - i \frac{\mu-\tilde\Pi}{p^2(1-\Pi)} \epsilon_{\mu\nu\alpha}p^\alpha\right) - i \xi \frac{p_\mu p\nu}{p^4}
\label{Dfull-Eq}
\ee
where the projector $P_{\mu\nu}$ is defined as

\be
P_{\mu\nu} = g_{\mu\nu} - \frac{p_\mu p_\nu}{p^2}
\ee
and the full photon self energy is parameterised as
\be
\Pi_{\mu\nu}(p) = p^2  P_{\mu\nu}\Pi(p) + i \epsilon_{\mu\nu\alpha}p^\alpha \tilde \Pi(p).
\ee

The precise statement of the Coleman-Hill theorem is

\be
\tilde\Pi(0) =   \alpha \sum_{i=1}^{N_f} \frac{m_i}{|m_i|}
\label{Pi2-Eq}
\ee
to all orders in perturbation theory. This quantity is therefore interpreted as the topological photon mass. We have introduced $\alpha = e^2/(4\pi)$. All subsequent numerical results are expressed in units of $\alpha$.

The proof of the theorem is relatively simple: 
if one considers the effective action obtained by integrating out fermions,
gauge invariance implies that the leading behaviour of a general $n$-point function is

\be
\Gamma^{(n)}(p_1 \ldots p_n) = O(p_1 p_2).
\ee
Two external photon lines  contribute $O(p^2)$ to any diagram beyond one-loop and so can not contribute to a function that is proportional to $p$. 
Nonzero contributions can only arise from graphs
in which the two external photon lines end on a loop that has no other photon lines attached to it. Thus all
corrections to the topological mass beyond one loop order vanish\footnote{The proof relies on a nonzero fermion mass. We believe that it remains true in the massless case as this limit must be taken after the computation. Nevertheless, there are counter claims, including computations with a nonzero photon mass\cite{ST}. These reveal $\tilde\Pi(0) \sim e^4/\mu$ at two-loop. However, this computation was made with explicitly massless fermions. This is the likely cause of the nonsensical result as $\mu\to 0$.}. 
It is also known that
the two-loop correction to $\tilde \Pi(0)$ vanishes\cite{KS}. The two results together suggest that corrections to the topological mass vanish order-by-order in perturbation theory. We will shortly develop expressions for $\Pi$ and $\tilde \Pi$ in the Schwinger-Dyson formalism. Since an arbitrary truncation of the Schwinger-Dyson equations sums diagrams in a way that need not be consistent with perturbation theory or gauge invariance, one cannot expect that the Coleman-Hill constraint on $\tilde\Pi(0)$ holds. Deviations from  Eq. \ref{Pi2-Eq} can then serve as a useful diagnostic for the accuracy of a given truncation scheme.

\begin{figure}[ht]
\includegraphics[width=10cm,angle=0]{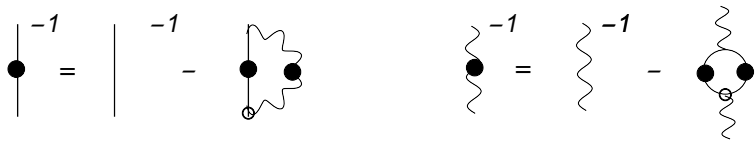}
\caption{Schwinger-Dyson Equations. Solid circles represent full propagators. The open circles represent a model vertex.}
\label{sde-fig}
\end{figure}

The Schwinger-Dyson equations for the two-point functions are shown in Fig. \ref{sde-fig}. Solutions to these equations yield the full photon propagator, parameterised in Eq. \ref{Dfull-Eq} and the full fermion propagator, defined by

\be
S(p) = \frac{i}{A(p)\rlap{/}p - B(p)}.
\ee
The Schwinger-Dyson equations have been truncated by assuming a model form for the vertex (denoted with a small open circle).  Typical model vertices include the rainbow ladder approximation:

\be
i\Gamma^\mu_{RL}(k,p) = \gamma^\mu,
\ee
the central Ball-Chiu vertex
\be
i\Gamma^\mu_{CBC}(k,p) = \frac{1}{2}(A(k)+A(p)) \gamma^\mu,
\ee
or the Ball-Chiu vertex

\be
i\Gamma^\mu_{BC}(k,p) = \frac{1}{2}\left(A(k)+A(p)\right)\gamma^\mu + \frac{1}{2}\frac{A(k)-A(p)}{k^2-p^2}(\rlap{/}k +\rlap{/}p)(k^\mu + p^\mu) - \frac{B(k)-B(p)}{k^2-p^2}(k^\mu + p^\mu) .
\ee
The Ball-Chiu vertex is the unique form of the longitudinal portion of the vertex that is consistent with the Ward-Takahashi identity and is free of kinematic singularities\cite{BC}. Use of the Ball-Chiu vertex is a necessary, but not sufficient, condition for gauge invariance of the solution.

The transverse portion of the full vertex remains unspecified. Curtis and Pennington have used multiplicative renormalisability to argue that the most important transverse term is\cite{CP}

\be
i\Gamma^\mu_{CP}(k,p) = \frac{1}{2} \frac{A(k)-A(p)}{d(k,p)} \left[ \gamma^\mu (k^2-p^2) - (k+p)^\mu \, (\rlap{/}k - \rlap{/}p)\right]
\ee
with 
\be
d(k,p) = \frac{(k^2-p^2)^2 + (M(k)^2+M(p)^2)^2}{k^2+p^2}
\label{d-eq}
\ee
where $M = B/A$ is the mass term of the full propagator. We shall refer to these truncations as RL (rainbow-ladder), CBC (central Ball-Chiu), BC (Ball-Chiu), or CP (following Ref. \cite{FADM}, this is the Ball-Chiu plus Curtis-Pennington vertex in the fermion propagator and the Ball-Chiu vertex in the photon propagator).

In rainbow-ladder approximation the Schwinger-Dyson equations for the $i$th fermion are

\be
B_i(p^2) = - 2 i e^2 \int\frac{d^3q}{(2\pi)^3} \, \left( \frac{B_i(q^2)(1-\Pi(K))}{(A_i^2q^2 - B_i^2)D(K)} 
 + \frac{(\mu-\tilde\Pi(K)) A_i(q^2) q\cdot K}{K^2\,(A_i^2q^2 - B_i^2)D(K)} 
 +  B_i(q^2)\frac{\xi}{2 K^2 (A_i^2 q^2 - B_i^2)}\right)
\label{B-Eq}
\ee
and

\begin{eqnarray}
A_i(p^2) &=& 1 - 2i \frac{e^2}{p^2}\int \frac{d^3q}{(2\pi)^3} \, \big[ A_i(q^2) \left( \frac{p\cdot K \,q\cdot K}{K^2}\,(1-\Pi(K)) + \xi \frac{D(K)}{2K^4}( p\cdot q \, K^2 - 2 K\cdot p K\cdot q)\right) + \nonumber \\
&+& B_i(q^2)\frac{p\cdot K(\mu-\tilde\Pi)}{K^2}\big] \frac{1}{(A_i^2q^2-B_i^2)D(K)}.
\label{A-Eq}
\end{eqnarray}
We have defined $K = p-q$ and 

\be
D(K) = K^2 (1 -\Pi(K))^2 - (\mu-\tilde\Pi(K))^2
\ee
in these expressions.

The rainbow-ladder expressions for the scalar photon functions are

\be
p^2 \Pi(p^2) = 2 i e^2 \sum_{i=1}^{N_f}\,\int \frac{d^3 q}{(2\pi)^3}\, \frac{A_i(q^2)\, A_i(Q^2)}{(A_i^2q^2-B_i^2)(A_i(Q)^2Q^2 - B_i(Q)^2)} \left(  q\cdot Q - 3 \frac{q\cdot p\,  Q\cdot p}{p^2}\right)
\ee
and
\be
p^2 \tilde \Pi(p^2) = -2 i e^2 \sum_{i=1}^{N_f}\,\int \frac{d^3 q}{(2\pi)^3}\, \frac{Q\cdot p\, B_i(q^2) A_i(Q^2) - q \cdot p A_i(q^2) B_i(Q^2)}{(A_i^2q^2-B_i^2)(A_i(Q)^2Q^2 - B_i(Q)^2)},
\label{FullPi2-Eq}
\ee
where $Q= p+q$.
The expressions with the BC or CP vertices are more complicated and not very illuminating and hence are given in the Appendix. 

Note that the anomalous photon function $\tilde\Pi$ arises in the equations for $A_i$ and $B_i$, thus hidden parity symmetry can affect chiral symmetry breaking and vice versa. Furthermore, (for  $\mu =0$) the relation
$\tilde\Pi[-B] = - \tilde\Pi[B]$ implies that the Schwinger-Dyson equations are invariant under $ B \to -B$. This remains true with the Ball-Chiu and Curtis-Pennington vertices.

Finally, computing perturbatively with dimensional regularisation yields the following results for the one-loop scalar photon functions:

\be
\Pi = -2 \alpha \sum_{i=1}^{N_f}\int_0^1 dx\, \frac{x(1-x)}{\sqrt{m_i^2 - x(1-x)p^2}}
\ee
and
\be
\tilde \Pi = \alpha \sum_{i=1}^{N_f}\int_0^1 dx \, \frac{m_i}{\sqrt{m_i^2 - x(1-x)p^2}}.
\ee
Thus $\Pi(0) = -\alpha/(3m) N_f$   and $\tilde\Pi(0)$ is given by Eq. \ref{Pi2-Eq}.
 We have confirmed that these results are recovered numerically when $A=1$ and $B=m$.

We shall consider two order parameters in the following, the dynamical mass

\be
M(p) = B(p)/A(p)
\ee
and the fermion condensate $ \langle \bar \psi \psi\rangle$.
Direct computation yields the Euclidean result

\be
\langle \bar \psi \psi\rangle = 2 \int \frac{d^3q_E}{(2\pi)^3} \frac{B}{A^2q_E^2+B^2}.
\ee
Alternatively, examination of Eq. \ref{B-Eq}  reveals
\be
B(p^2) \to 2\pi\alpha\frac{2+\xi}{p_E^2} \langle \bar \psi \psi\rangle.
\label{cond-eq}
\ee
We stress that this result is true regardless of the vertex approximations made. Its general validity derives from the operator product expansion\cite{pol}.

As mentioned in the introduction, Appelquist {\it et al.} have analysed the Schwinger-Dyson equations in the large $N_f$ limit and obtained a critical number of fermion flavours, $N_\star \approx  N_\star^A = 64/\pi^2$ for two-component fermions, above which chiral symmetry is restored. Of course, one can criticise this computation for deriving a relatively small value of $N_\star$ in the large $N_f$ limit.  More recent versions of this calculation are made on Refs. \cite{maris} and \cite{FADM}. The result is of the form

\be
B(0) \propto N_f\, \exp\left(\frac{-2\pi}{\sqrt{N_\star^A/N_f - 1}}\right).
\label{Sigma-Eq}
\ee
As mentioned, this result is not entirely accepted; we shall analyse its accuracy numerically in the next section.

Finally, Pennington and Walsh have used renormalisation group arguments to suggest that the wavefunction renormalisation behaves as\cite{PW}

\be
A(p) \sim \left(\frac{p^2+m^2}{e^2}\right)^{4/3\pi^2 N_f}.
\ee
We consider the case $m=0$ so this equation might imply that $A$ runs to zero with momentum. However the dynamical fermion mass is generically expected to mimic the bare fermion mass, and one can anticipate that $A$ is finite at the origin but approaches zero as $N_f$ is increased past $N_\star$.

Finally, we choose to work in Landau gauge, which is widely regarded as providing the most reliable results when working at this level of truncation. For detailed discussions of gauge variance we refer the reader to Ref. \cite{gg}.

\section{Numerics and Results}

\subsection{Numerical Methods}

The generic numerical problem we seek to solve is of the form
$X = F[X]$
where $X$ represents the unknown functions $A$ and $B$ and $F$ is a functional denoting the nonlinear integrals on the right hand side of Eqs. \ref{B-Eq},\ref{A-Eq}. Notice that the photon functions are relationships, not equations to be solved. We employed a number of strategies to solve these equations, including:

(i) natural iteration. 

\be
X^{(n+1)} = F[X^{(n)}]
\ee

(ii) implicit iteration

\be
X^{(n)} = F[X^{(n+1)}]
\ee

(iii) iteration with over-relaxation

\be
X^{(n+1)} = (1-\omega) X^{(n)} + \omega F[X^{(n)}]
\ee

(iv) Newton-Raphson iteration

\be
X^{(n+1)} = (1 - \frac{\delta F}{\delta X}[X^{(n)}])^{-1}(F[X^{(n)}] - X^{(n)})
\ee

(v) minimisation

\be
{\rm min}\, || X - F[X] ||.
\ee

In addition, a variety of representations of the functions $A$ and $B$ were attempted including the point basis, modified Legendre polynomials, and a quadratic+inverse quadratic spline. We found that natural iteration worked quite well for all vertices except Ball-Chiu. Of course simply iterating need not yield a solution at all, rather the algorithm could approach a limit cycle, or could even go through a period doubling transition to a chaotic regime. 
Newton-Raphson iteration is both faster and more stable (but not always) than natural iteration. 
Over-relaxation works quite well if Gauss-Seidel updates are used. As with elliptic partial differential equations, one must tune the value of $\omega$ carefully.
We did not find cases where implicit iteration was useful. Minimisation is always stable, but can be very slow, especially when a large number of degrees of freedom are being varied. Unfortunately, the method can easily converge to one of a great many local minima. Worse, these minima need not be close to an actual solution to the Schwinger-Dyson equations and care is required.

\subsection{Results}

\subsubsection{Parity Symmetry}

We seek to confirm the reflection symmetry breaking pattern suggested by Appelquist {\it et al.}. Thus if parity symmetry is maintained one expects either no chiral symmetry breaking or chiral symmetry breaking with $N_f/2$ fermions of mass $M$ and $N_f/2$ fermions of mass $-M$. If reflection symmetry is broken then
one expects $N_f$ fermions with mass $M$ (or $-M$). (Other patterns are possible; we restrict attention to these three).

We have performed computations assuming a reflection symmetry breaking pattern of ($M$,$M$) with all vertex models.
 The natural iteration algorithm approaches a limit cycle of the type $(A,B) \to (A,-B)$. While this could prove to be a nontrivial solution (due to the $B$ reflection symmetry discussed above), the error $||X-F[X]||$ does not decrease with iteration. This suggests 
that the limit cycle does not represent an actual solution to the Schwinger-Dyson equations. Of course, the lack of convergence of a natural iteration algorithm does not prove that a solution does not exist. We have therefore repeated the computation with the minimisation algorithm. In this case $A$ evolves to a smooth function while $B$ approaches zero (with substantial noise). Similar behaviour was seen in all vertex models. Thus it appears that a nontrivial reflection-breaking solution to the Schwinger-Dyson equations 
does not exist, as suggested in Refs.\cite{Appelquist-parity}, \cite{stam}. This computation is the first extension of those conclusions to the full two-point function Schwinger-Dyson equations with vertex models of varying sophistication.

Recall that Ref. \cite{Hoshino} found a parity-violating solution. Their computation was made in the quenched rainbow-ladder approximation to the Schwinger-Dyson equations. We, in fact, confirm their result; however, it is clear that the quenched approximation is unsuitable for investigating properties of QED3. The photon propagator is changed drastically in the infrared, even in perturbation theory, and this must be accounted for in the formalism.

\subsubsection{The Coleman-Hill Constraint}

\begin{figure}[ht]
\input{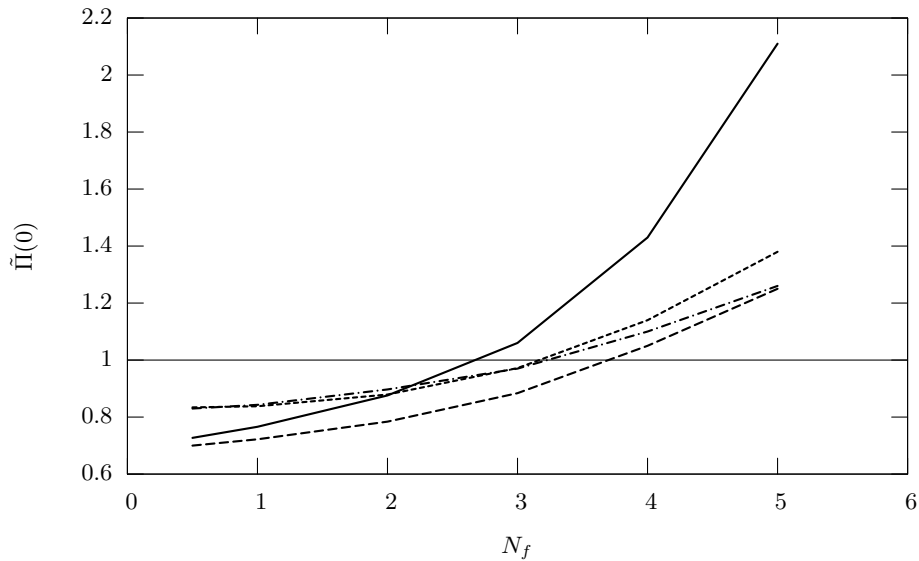}
\caption{$\tilde \Pi(0)$ vs. $N_f$ for various models. From top to bottom at $N_f=5$ these are: RL, CBC, CP, BC.}
\label{pi2-fig}
\end{figure}

We have argued that the value of $\tilde \Pi(0)$ can be used to test the accuracy of truncation schemes in the Schwinger-Dyson equations because its value is fixed by Eq. \ref{Pi2-Eq} to all orders. We display 
$\tilde \Pi(0)$ as a function of the number of fermions for the RL, CBC, BC, and CP Ans{\"a}tze in Fig. \ref{pi2-fig}. As hoped, more sophisticated vertex Ans{\"a}tze maintain the Coleman-Hill result more accurately and over a larger range of $N_f$. The CP vertex remains within 20\% of unity over the range of $N_f$ considered. This is a highly nontrivial phenomenon: as $N_f$ increases $B$ becomes exponentially small, driving $\tilde \Pi$ to also be exponentially small (see Eq. \ref{FullPi2-Eq}). This is countered by an increasingly exposed infrared divergence in the integral. It is remarkable that the two effects cancel so precisely when the CBC, BC, or CP  Ans{\"a}tze are employed. Because of this, we estimate that the truncation error in our CP results is less than 20\% over the range of $N_f$ considered here.

\subsubsection{Chiral Symmetry}

\begin{figure}[ht]
\qquad
\includegraphics[width=5.5cm,angle=270]{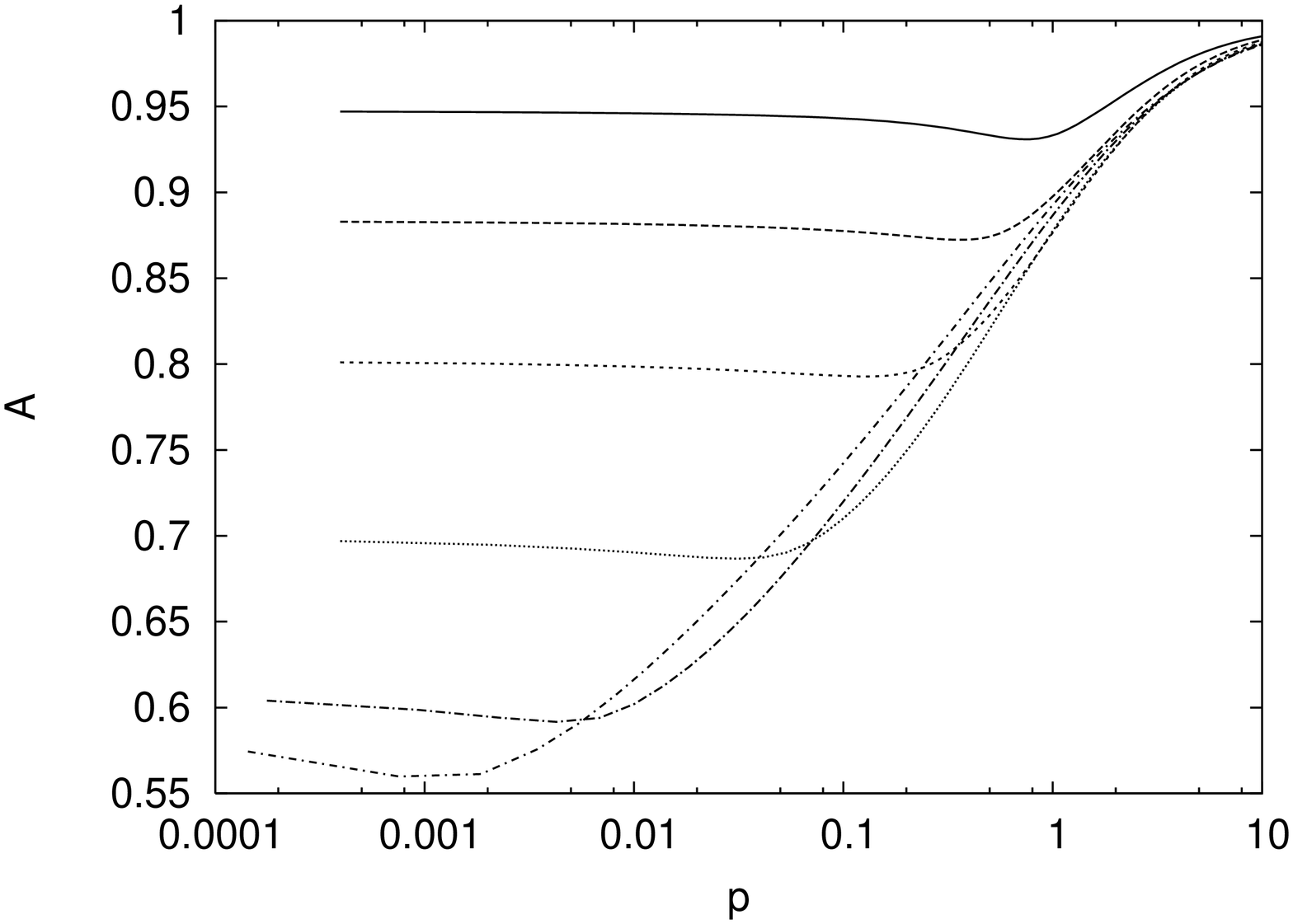}
\qquad
\includegraphics[width=5.5cm,angle=270]{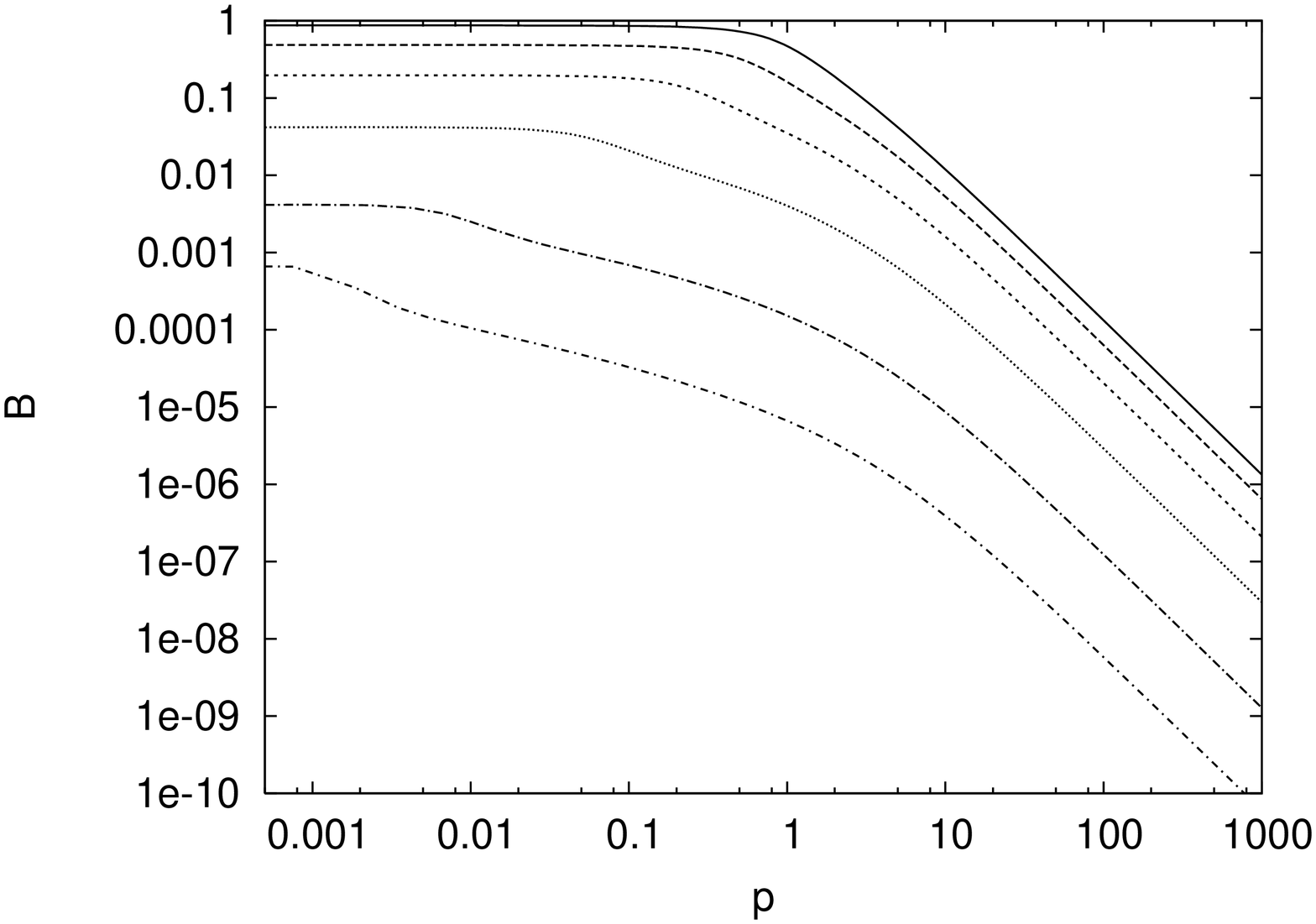}
\caption{$A$ vs. $p$ for various $N_f$ (L). $B$ vs. $p$ for various $N_f$ (R). These curves are obtained in the CBC truncation and correspond to $N_f$ = 1, 2, 3, 4, 5, and 5.75 from top to bottom.}
\label{B-fig}
\end{figure}

\begin{figure}[ht]
\includegraphics[width=5.5cm,angle=270]{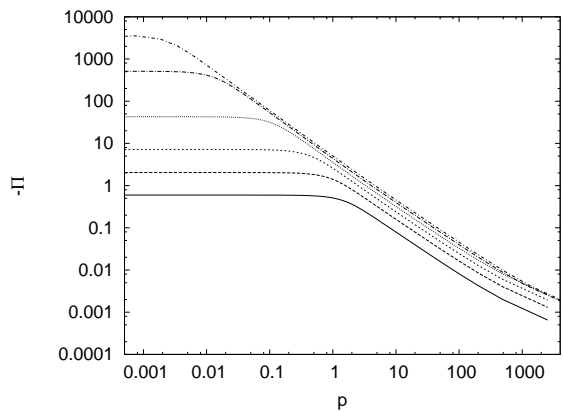}
\caption{Vacuum Polarisation vs. momentum for various $N_f$. These curves are obtained in the CBC truncation and correspond to $N_f$ = 1, 2, 3, 4, 5, and 5.75 from bottom to top.}
\label{Pi-fig}
\end{figure}

As expected, we find chiral symmetry breaking solutions for small $N_f$ for all vertex models. The $N_f$-behaviour of the solutions is explored in Figs. \ref{B-fig} and \ref{Pi-fig}. 
One sees that $B(p)$ drops precipitously with increasing number of fermion flavours. Whether $B$ vanishes at a finite value of $N_f$ will be discussed below. First we note that the value of $A$ drops near the origin and may be approaching a limit function with $A(0)=0$. Does this happen at a finite $N_f$? A similar behaviour is seen for $\Pi$, again, with the possibility that an infrared singular limit function is approached at a finite value of $N_f$. These figures have been generated with the CBC vertex, qualitatively, and in most parts quantitatively, similar results are obtained with the other truncations.

\begin{figure}[ht]
\includegraphics[width=6cm,angle=270]{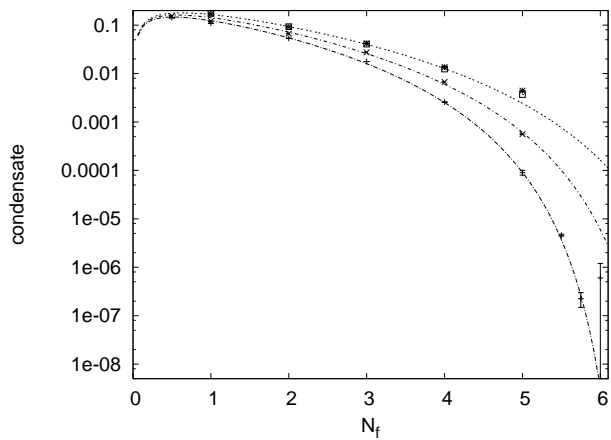}
\caption{The condensate for (top to bottom) BC, CP, RL, and CBC models, with fit functions (described in the text).}
\label{cond-fig}
\end{figure}

We examine the putative phase transition more closely by plotting the fermion condensate  versus $N_f$ (Fig. \ref{cond-fig}). We have confirmed that the directly measured condensate agrees with that obtained from Eq. \ref{cond-eq} to high accuracy.
Results for the RL, CBC, BC, and CP models are shown in Fig. \ref{cond-fig}. Also shown are fits with the Ansatz

\be
\langle \bar \psi \psi\rangle(N_f) = a N_f \exp\Bigg(\frac{-2\pi}{\sqrt{N_\star/N_f - 1}}\Bigg).
\label{cond-fit-eq}
\ee

It is quite difficult to obtain these results for large values of $N_f$:  extrapolation in the number of integration points is necessary for $N_f > 4$.  Our most accurate results are for the CBC vertex and are in excellent agreement with the Ansatz of Eq. \ref{cond-fit-eq}, implying strongly that a chiral restoration phase transition does indeed occur. An Ansatz like $\langle \bar \psi \psi\rangle \propto \exp(-N_f^4)$ also does a reasonable job fitting the CBC points, and of course does not exhibit a phase transition. Nevertheless the quality of this fit is significantly lower than that of Eq. \ref{cond-fit-eq}. 

Remarkably, a simple adjustment of $N_\star$ in Eq. \ref{cond-fit-eq} provides excellent fits to the RL, BC, and CP data as well. We obtain $a=1.8$ and

\be
N_\star({\rm CBC}) = 1.00 \cdot N_\star^A
\ee
\be
N_\star({\rm RL}) = 1.10 \cdot N_\star^A
\ee
\be
N_\star({\rm CP}) = N_\star({\rm BC}) = 1.21 \cdot N_\star^A.
\ee
Although we can be less certain that the BC and CP models exhibit symmetry restoration, the good fits, the similarity of all model results, and the theoretical motivation all make this very likely in our opinion. Our numerical results appear to be in agreement with earlier results \cite{maris, FADM}.

\section{Conclusions}

Previous arguments for the lack of reflection symmetry breaking in two-component QED3 relied on the large $N_f$ expansion and several simplifying assumptions. We have extended the analysis to  finite $N_f$ with several model vertices, including the Ball-Chiu-Curtis-Pennington model, which is the most sophisticated model of QED3 to date. We cannot find nontrivial solutions for the fermion propagator with several numerical techniques, making it very likely that parity symmetry is maintained in two-component QED3.

A novel feature of this analysis has been the use of the Coleman-Hill theorem as a diagnostic for truncation accuracy. General arguments imply that truncations that obey the Ward-Takahashi identity should perform better than those that do not. We have confirmed this. We observe that deviations from the exact value of $\tilde\Pi(0)$ indicate truncation errors of approximately 20\% over a large range of $N_f$, so there is some hope that the conclusions obtained here are of reasonable fidelity.

Parity-symmetric two-component QED3 is equivalent to four-component QED3, permitting us to examine the issue of chiral symmetry breaking and restoration. 
The difference between symmetry restoration and exponentially suppressed symmetry breaking is necessarily small, and numerically distinguishing these scenarios can be very challenging. 
While not definitive, our results agree with earlier computations in the Schwinger-Dyson formalism and hence support the original simple arguments of Appelquist {\it et al.}. Furthermore, it is likely that the transition occurs in all the model truncations considered here. The value of the critical number of fermions varies by model, but all are surprisingly close to the original estimate.

\acknowledgments
This research was supported by the U.S. Department of Energy under contract
DE-FG02-00ER41135 and an Andrew W. Mellon Predoctoral Fellowship.

\section{appendix}

The truncated Schwinger-Dyson equations for the fermion propagator with the BC+CP vertex are as follows.

\be
B(p) = m - i e^2 \int \frac{d^3q}{(2\pi)^3}\, \left[ \frac{\bar A+d2}{D} \beta_1 + \frac{\Delta A}{D} \beta_2 + \frac{\Delta B}{D} \beta_3 + \frac{d2}{D(q^2-p^2)} \beta_4 \right]
\ee
and
\be
A(p) = 1 + i \frac{e^2}{p^2} \int \frac{d^3 q}{(2\pi)^3} \, \left[ \frac{\bar A+d2}{D} \alpha_1 + \frac{\Delta A}{D} \alpha_2 + \frac{\Delta B}{D} \alpha_3 + \frac{d2}{D(q^2-p^2)} \alpha_4 \right].
\ee
Subexpressions are
\be
\bar A = \frac{1}{2}(A(p)+A(q))
\ee
\be
\Delta A = \frac{1}{2} \frac{A(q)-A(p)}{q^2-p^2}
\ee
\be
\Delta B = -\frac{B(q)-B(p)}{q^2-p^2}
\ee
\be
D = A(q)^2 q^2 - B(q)^2
\ee
\be
d2(p,q) = \frac{1}{2} \frac{(A(q)-A(p))(q^2-p^2)}{d(q,p)}
\ee
See Eq. \ref{d-eq} for $d(q,p)$. 

The coefficients are 
\be
\beta_1 = 2aB - 2 A b q\cdot K + \xi B/K^2
\ee
\be
\beta_2 = a B(Q^2 - Q\cdot \hat K Q\cdot \hat K) + \xi B (Q\cdot K)^2/K^4 - A b (p\cdot Q q \cdot K - p\cdot K q\cdot Q)
\ee
\be
\beta_3 = A a (Q\cdot q - Q\cdot \hat K q \cdot \hat K) + \xi A Q\cdot K q\cdot K/K^4
\ee
\be
\beta_4 = -\xi B (p^2-q^2)/K^2 - 2 A b (p^2q^2 - (p\cdot q)^2)
\ee
\be
\alpha_1 = -2 A a p\cdot \hat K q\cdot \hat K + \xi A (2p\cdot K q\cdot K - p\cdot q K^2)/K^4 + 2 B b p\cdot K
\ee
\be
\alpha_2 = \frac{1}{2} A a (q^2+p^2) (Q^2 - (Q\cdot \hat K)^2) + \xi A (Q\cdot K)^2 p\cdot q/K^4 - B b (p\cdot Q q\cdot K - p\cdot K q\cdot Q)
\ee
\be
\alpha_3 = - A b (p\cdot Q q \cdot K - p\cdot K q\cdot Q) + a B(p\cdot Q - p\cdot\hat K Q\cdot \hat K) + \xi B p\cdot K Q\cdot K/K^4
\ee
\begin{eqnarray}
\alpha_4 &=& -2 B b (p^2q^2 - (p\cdot q)^2) + A \Bigg[
 -a [p\cdot K (q\cdot Q-q\cdot K (p^2-q^2)/K^2) + \nonumber \\
&+& 
     q\cdot K (p\cdot Q-p \cdot K (p^2-q^2)/K^2)] - \nonumber \\
   &&   \xi(2 p\cdot K q\cdot K (p^2-q^2) - K^2p\cdot q(p^2-q^2))/K^4\Bigg]
\end{eqnarray}

\be
a = \frac{1-\Pi(K)}{K^2(1-\Pi(K))^2 - \tilde \Pi(K)^2}
\ee
\be
b = -\frac{\tilde \Pi(K)}{K^2(1-\Pi(K))} \cdot a
\ee
\be 
K^\mu = p^\mu - q^\mu \qquad Q^\mu = p^\mu + q^\mu
\ee

Ball-Chiu expressions for the photon scalar functions are
\be
p^2 \Pi(p)= 2 i e^2 \int \frac{d^3 q}{(2\pi)^3}\, \left[ \frac{\bar A}{D} \pi_1 + \frac{\Delta A}{D} \pi_2 + \frac{\Delta B}{D} \pi_3\right]
\ee
and
\be
p^2 \tilde \Pi(p) = - 2 i e^2 \int \frac{d^3q}{(2\pi)^3} \, \left[ \frac{\bar A}{D} \tilde \pi_1 + \frac{\Delta A}{D} \tilde \pi_2 + \frac{\Delta B}{D} \tilde \pi_3\right].
\ee
In this case
\be
D = [A(q)^2 q^2 -B(q)^2]\,[A(Q)^2 Q^2 - B(Q)^2]
\ee
and the subexpressions are given by (note that the arguments of $\bar A$, $\Delta A$ and $\Delta B$ are now $Q$ and $q$)
\be
\pi_1 = A(q)A(Q) (q\cdot Q - 3q\cdot \hat p Q \cdot \hat p)
\ee
\begin{eqnarray}
2 \pi_2 &=& A(q)A(Q)\, [2 q\cdot (q+Q) Q\cdot (Q+q) - (Q+q)^2 q\cdot Q - 3 q\cdot \hat p (Q+q)\cdot \hat p Q\cdot (Q+q) + \nonumber \\
&+& 3 (Q+q)\cdot \hat p (Q+q)\cdot \hat p q\cdot Q - 3 Q\cdot \hat p (Q+q)\cdot \hat p q\cdot (Q+q)] + \nonumber \\
&+& B(q)B(Q) [ (Q+q)^2 - 3 (Q+q)\cdot \hat p (Q+q)\cdot \hat p]
\end{eqnarray}
\be
2 \pi_3 = A(q) B(Q) (q\cdot (q+Q) - 3 q\cdot \hat p (Q+q)\cdot \hat p) +
B(q)A(Q)(Q\cdot (Q+q) - 3 Q\cdot \hat p (Q+q)\cdot \hat p
\ee
\be
\tilde\pi_1 = B(q)A(Q) p\cdot Q - A(q) B(Q) p\cdot q
\ee
\be
\tilde \pi_2 = (A(q)B(Q) + A(Q)B(q))\cdot (q^2p^2- (q\cdot p)^2)
\ee
\be
2 \tilde \pi_3 = A(q)A(Q)(q\cdot (Q+q) p^2 - p\cdot (Q+q) p\cdot q)
\ee


\end{document}